\documentclass[a4paper,preprint,showpacs,amssymb,superscriptaddress]{revtex4}

\usepackage{graphicx}

\begin{document}

\title{Diffusion in Fluctuating Media: The Resonant Activation
Problem}

\author{Jorge A. Revelli$^{a}$, Carlos. E. Budde$^{b}$
and Horacio S. Wio$^{a, c}$}

\affiliation{(a) Grupo de F\'{\i}sica Estad\'{\i}stica, Centro
At\'omico Bariloche and
Instituto Balseiro, \\ 8400 San Carlos de Bariloche, Argentina; \\
(b) Facultad de Matem\'aticas, Astronom\'{\i}a y F\'{\i}sica,
Universidad Nacional de C\'ordoba
\\ 5000 C\'ordoba, Argentina; \\ (c)Departament de F\'{\i}sica,
Universitat de les Illes Balears and IMEDEA \\ E-07122 Palma de
Mallorca, Spain}



\begin{abstract}
We present a one-dimensional model for diffusion in a fluctuating
lattice; that is a lattice which can be in two or more states.
Transitions between the lattice states are induced by a
combination of two processes: one periodic deterministic and the other
stochastic. We study the dynamics of a system of particles
moving in that medium, and characterize the problem from different
points of view: mean first passage time (MFPT), probability of
return to a given site ($P_{s_0}$), and the total length
displacement or number of visited lattice sites ($\Lambda$). We
observe a double {\it resonant activation}-like phenomenon when we
plot the MFPT and $P_{s_0}$ as functions of the intensity of the
transition rate stochastic component.
\end{abstract}


\maketitle

\vskip 1.truecm
\newpage
\normalsize

\subsection{Introduction}

The problem of diffusion in media submitted to global and/or local
fluctuations has strongly attracted the attention of researchers.
The main motivation was its interest in the study and analysis of
a large variety of processes. To name just a few examples, we
consider: random-walks in disordered systems \cite{I1,I2,I3}
(ionic conduction in polymeric solid electrolytes \cite{I4});
transport of Brownian particles that can be in two or more states
executing a diffusion process in each of them  but with different
diffusion constants \cite{I5,I6}; resonant activation over
fluctuating barriers \cite{I7} and escape from fluctuating systems
\cite{I8,I8P}; diffusion of ligand with stochastic gating
\cite{I9,I10,I91}; dynamical trapping problems
\cite{I11,I12,I140,I122}. Examples of potential applications are
charge transport in molecular scale systems \cite{I1b} and
photochemical reactions \cite{I2b} among others.

The above indicated phenomena share the property that the
switching between the different configurations or states of the
medium is independent of the transport or diffusion processes of
the walker. Usually, it is assumed that the states are independent
of each other and that the particles are subject to a Markovian
process inducing their motions within each state, described by
Master or Fokker-Planck equations.

In previous studies we have considered on one hand the problem of
diffusion in fluctuating media subject to Markovian and/or non
Markovian switchings \cite{nos1}. On the other hand we have
studied a {\it stochastic resonance}-like phenomenon in a gated
trapping system \cite{I122}. Motivated by those previous works, we
study here the evolution of particles diffusing in a fluctuating
medium when the switching mechanism that governs the transition
between the states has two components: one deterministic and the
other stochastic. The parameters were chosen in such a way that
the effect of the deterministic mechanism alone is not enough to
produce the switching. However, the simultaneous action of both
signals can induce the jump over the potential barrier and produce
the switching in the medium. Here, we consider a periodic signal
for the deterministic process, while the stochastic component is
characterized by a Gaussian white noise of zero mean and intensity
$\xi_0$. In this way we expect to capture some basic aspects of
the dynamics of more realistic diffusing systems and find
situations where the tuning between the stochastic and the
deterministic rates can induce some resonant phenomena as in
\cite{I122}.

In order to study this problem we have analyzed the {\it Mean
First Passage Time} (MFPT) in an infinite system. We have also
analyzed the probability that the particle returns to the initial
position ($P_{s_0}$), characterizing in this way the system's
capacity to restore its initial condition. We also study a measure
of the total length displacement or number of distinct visited
lattice sites. As it is well known, the MFPT is an extremely
important quantity because it gives insight into the reaction
processes (trapping, etc) and their dependence on the details of
the underlying dynamics of the systems. The MFPT also provides a
useful measure of the efficiency of the trapping and has been
investigated in a broad range of problems from chemistry to
biology \cite{I3b,I3c,I4b,I5b}.

The main goal of this work is the observation of a double {\it
resonant activation}-like phenomenon when we analyze the MFPT and
the $P_{s_0}$ as functions of the noise intensity of the
stochastic component of the transition rate among lattice states.

The organization of the paper is as follows. In the next Section
we describe the model in a formal way. After that, we present the
results of numerical simulations. In the final Section we discuss
the results and draw some conclusions.

\subsection{The Model}

In order to fix ideas we start considering the problem of a
particle performing a random walk on a continuous or discrete
fluctuating medium characterized by $N$ states which are labelled
by the index $j$ (with $j=1,2,...N)$. Following van Kampen
\cite{I13} we define the probability $u_j(t)$ that the medium has
stayed in the $j$ state after a time $t$ since its arrival at
$t=0$ as
\begin{equation}
\label{modelu}
u_{j}(t) = \exp\left(-\int_{0}^{t} \sum_i
\gamma_{ij}(t')dt' \right),
\end{equation}
where $\gamma_{ij}(t)$ is the probability per unit time for the
medium to jump from level $j$ to level $i$ and $t$ is the time it
has sojourned in $j$.

The {\it switching statistics} of the medium $v_{ij}(t)$; defined
as the probability that the medium ends its sojourn in the state
$j$ after a time between $t$ and $t+dt$ since it arrived at state
$j$ at $t=0$ by jumping to a given state $i$; is
\begin{equation}
\label{model1}
v_{ij}(t) dt = u_j(t) \gamma_{ij}(t) dt.
\end{equation}
It is worth remarking here that Eq. (\ref{model1}) is completely
general and no extra assumption has been made in writing it
\cite{I13}. As is well known, if the $v_{ij}(t)$ are exponential
functions of time, that is the $\gamma_{ij}(t)$ are
$t$-independent functions (see for instance \cite{I3b}), we have a
Markovian switching process between the states of the medium.

In order to simplify the problem, we assume that the system can
fluctuate between only two states. The new aspect here is that we
have used a combination of noisy and deterministic switching
mechanisms to describe the transitions between those states. We
assume that $v_{ij}(t)$ is given by
\begin{equation}
\label{modelv}
v_{ij}(t) = \theta[B \sin (\omega
t)+\xi-\xi_c],
\end{equation}
where $\theta (x)$ is the step function, and determines when the
lattice is in one state or in the other. The dynamics of this
mechanism is the following: if the signal, composed of a harmonic
part plus the noise contribution $\xi$, reaches a threshold
$\xi_c$ the lattice changes its state, otherwise it doesn't.  We
are interested in the case where $\xi_c>B$, that is, without noise
the deterministic signal is not able to induce a change of the
lattice ´s state.

Finally, in order to complete the model, we must give the
statistical properties of the noise $\xi$. We assume that $\xi$ is
an uncorrelated Gaussian noise of width $\xi_0$, i.e.
\begin{equation}
\label{noise}
\langle \xi (t) \xi (u) \rangle= \delta_{t,u} \xi_0^2.
\end{equation}
Note that this is not a standard white noise \cite{I3c} due to the
Kronecker symbol (instead of Dirac $'s$ delta $\delta(t-u)$). For
each fixed state $j$ of the medium  the transport process is
Markovian and we denote its corresponding ``propagator" \cite{I13}
by $A_j$. These propagators are differential operators (matrices)
in the case of a continuous (discrete) medium and its structure
 depends strongly on the character of the fluctuations.

In \cite{nos1} we have been able to obtain some analytical results
for the MFPT in the case of Markovian and non-Markovian processes
that were satisfactorily compared with numerical simulations.
However, and due to the complex dynamics of the present case, we
were not able to solve this problem analytically and have resorted
to MonteCarlo simulations.

The numerical simulations were performed on a one dimensional
infinite lattice. The particles were initially located at the
lattice site $s_0$ in a given state. We assumed that the
propagators $A_j$ were arrays $W$ describing a discrete one
dimensional random walk in each state of the lattice, and also
assumed jumps only to first neighbors. The jumps within each state
were characterized by two parameters: $\lambda_j$, the temporal
rate of jump and $\eta_j$, the ``bias"  to make a jump in a given
direction. For the case in which we studied the {\it return to the
origin}, the system started its motion at the origin. All
simulations shown in the following figures correspond to averages
over $100000$ realizations.

In the next Section we present several figures where we show the
results of our simulations. In the whole study we have used state
$1$ as the standard state, characterized by the following
parameters: $\lambda_1 = 1$ and $\eta_1 = 1$ (we have chosen the
bias to the left side of the lattice), while state $2$ is the test
state characterized by $\lambda_2 = 1$.We have chosen the same
temporal rates because we want to focus on the effect of the bias
and the noise intensity on the system response. Hence, $\eta_2$
together with $\xi_0$, are the parameters we varied. In addition,
and for all figures, we have adopted the following parameters for
the deterministic signal: frequency $\omega = 1.$, and amplitude
$B = 1.$; while the activation threshold is $\xi_c = 2$.

\subsection{MonteCarlo Results}

\begin{figure}
\centering
\resizebox{.6\columnwidth}{!}{\includegraphics{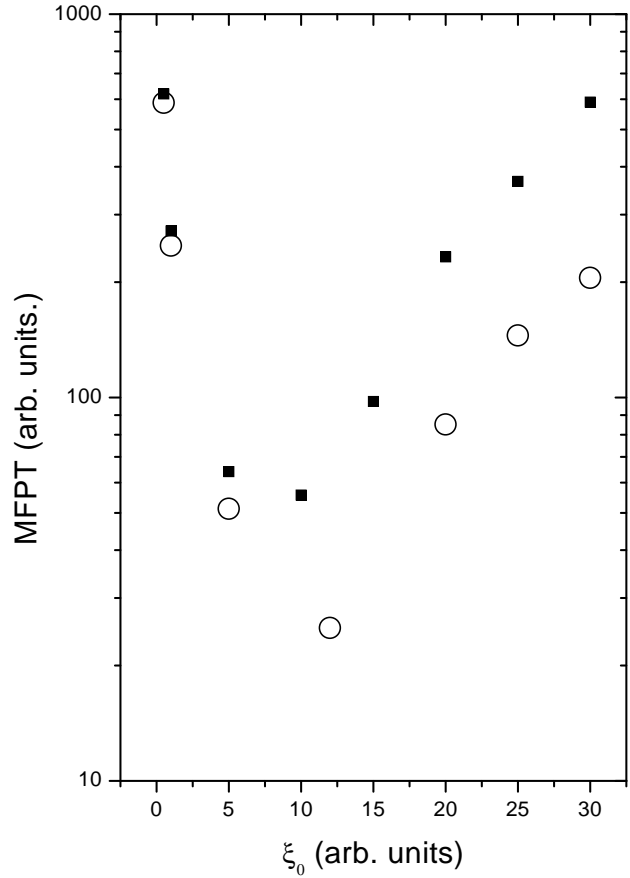}}
\caption{MFPT vs $\xi_0$, for $\eta_2 = 0.1$(open circles) and
$\eta_2 = 0.05$ (black-squares).} \label{f14}
\end{figure}

In Fig. \ref{f14} we show the results for the MFPT to reach the
origin ($s = 0$) for a particle that started its motion at site
$s_0 = 4$, as a function of the noise intensity $\xi_0$. From this
figure, it is apparent that there is an optimal value $\xi_0$ for
which the MFPT has a minimum (or viceversa, the ``activation rate"
has a maximum). This behavior, that is a manifestation of the so
called {\it resonant activation} phenomenon \cite{I7,I8,I8P}, is
one of the main results of this work. When $\xi_0$  is small, the
transition rate is low, and the particle is in an ``unfavorable"
state, it will remain there for a long time, contributing to a
larger MFPT. When $\xi_0$ is large, the transition rate will be
high, and the particle will not show a net motion, again yielding
a large MFPT. But there is an optimal $\xi_0$ where the
combination between transition rate and motion yields a minimum of
the MFPT. It is worth remarking here that this situation occurs
when the bias in each state is strong enough but they are in opposite
directions, however the phenomenon disappears when the tendencies
in both states are similar.

\begin{figure}
\centering
\resizebox{.6\columnwidth}{!}{\includegraphics{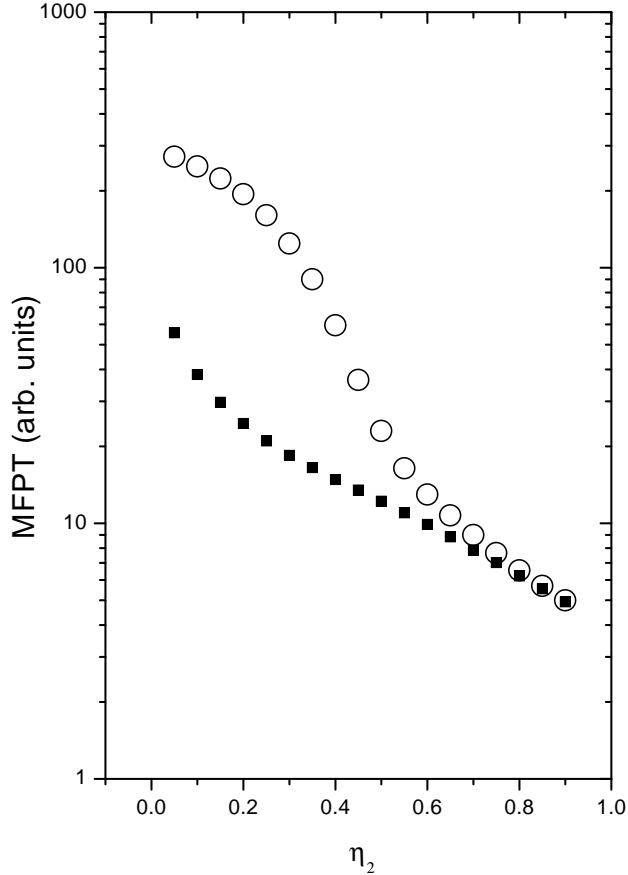}}
\caption{MFPT vs $\eta_2$, the bias in state $2$. The circles
corresponds to the MFPT values for $\xi_0=1.0$, and the squares
for $\xi_0 = 10.0$.} \label{f24}
\end{figure}

Figure \ref{f24} shows the dependence of the MFPT on $\eta_2$ for
two values of $\xi_0$ corresponding to low and high noise
intensities. In this figure we can recognize two main aspects.
Firstly, the MFPT decreases as $\eta_2$ grows, which is a logical
result because increasing $\eta_2$ means that we have an
increasing bias to the origin. Secondly the figure shows an
important difference in the MFPT for small $\eta_2$, where we see
that the noise intensity plays an important role. For small
$\xi_0$ the transition rate between the states is low and the
particle can remain a long time moving to or away from the origin.
For large $\eta_2$, the differences between the behavior for
different noise intensities disappear. Also, at intermediate
$\eta_2$ values, the MFPT remains higher for a smaller noise
intensity.

\begin{figure}
\centering
\resizebox{.6\columnwidth}{!}{\includegraphics{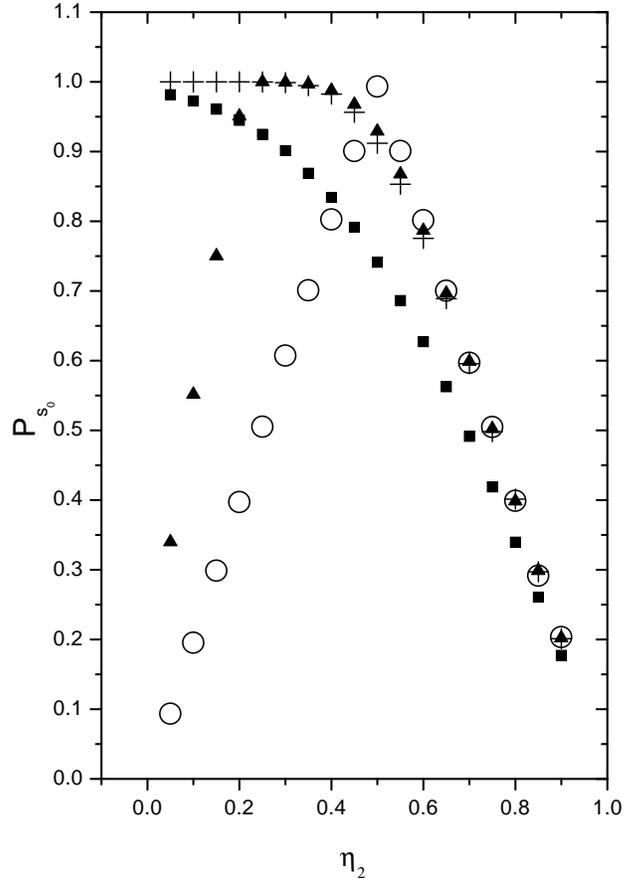}}
\caption{$P_{s_0}$ vs $\eta_2$. The circles represents the
one-state case, the squares are the data for $\xi_0 = 5.0$, the
crosses for $\xi_0 = 10.0$ and the triangles for $\xi_0 = 20.0$.}
\label{f34}
\end{figure}

Figure \ref{f34} depicts $P_{s_0}$, the probability of return to
the origin, as a function of $\eta_2$. The initial condition of
the system was at site zero ($s_0=0$) in state one, and the
simulation time was $10 000$. For comparison, the case of only one
lattice is also shown in this figure. The latter presents a
maximum for an intermediate value of $\eta_2$, that is when there
is no privileged direction, otherwise the probability decreases as
the particle is forced to move away from the origin. For the two
state case, we can see that, for small noise intensities, the
noise enhances the system {\it response} for a fixed (small) value
of $\eta_2$. This is due to the possibility of changing from the
unfavorable state to the one where the particle has a larger
probability of moving towards the origin. If we increase the noise
intensity, the response is enhanced, that is the return
probability becomes larger. However, if we continue increasing the
noise intensity finally the response for low $\eta_2$ is reduced
and there appears a maximum in the probability indicating that
there is an {\it optimal} value of $\eta_2$. For large $\eta_2$
values, the response becomes smaller for all noise intensities as
expected. Here we have another important result of our study, a
kind of {\it double resonant} behavior both in $\xi_0$ and
$\eta_2$.

\begin{figure}
\centering
\resizebox{.6\columnwidth}{!}{\includegraphics{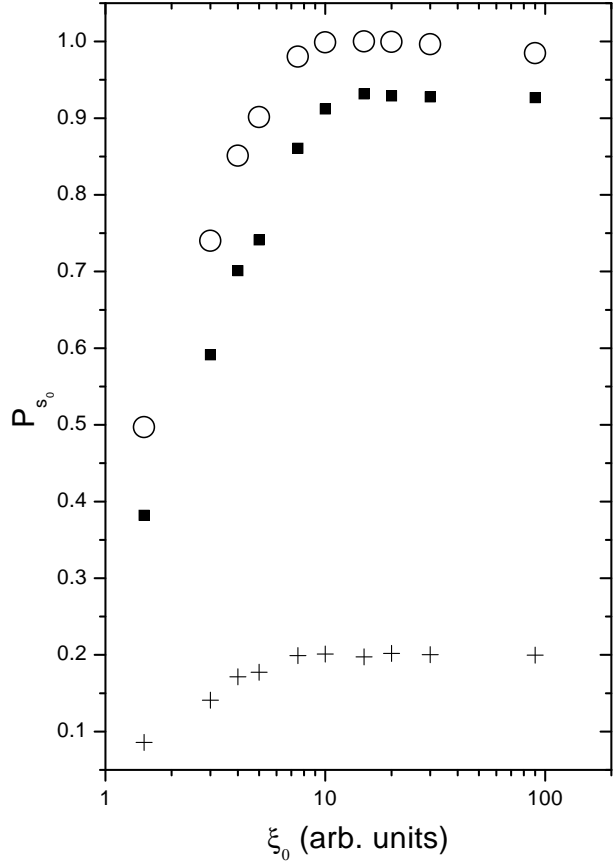}}
\caption{$P_{s_0}$ vs $\xi_0$. The circles represents the data
for $\eta_2 = 0.3$, the squares for $\eta_2 = 0.5$ and the crosses
for $\eta_2 = 0.9$} \label{f44}
\end{figure}

Figure \ref{f44} depicts $P_{s_0}$ as a function of the noise
intensity $\xi_0$, for different values of $\eta_2$. The increase
of this probability with increasing $\xi_0$ is apparent, reaching
a kind of plateau for $\xi_0 \geq 10$. However, for $\eta_2 =
0.3$, a maximum around $\xi_0 = 10$ is insinuated, resembling a
resonant like phenomenon. The fact that this probability decreases
with increasing $\eta_2$ is again a logical result because larger
$\eta_2$ implies an increasing bias to the origin.

\begin{figure}
\centering
\resizebox{.6\columnwidth}{!}{\includegraphics{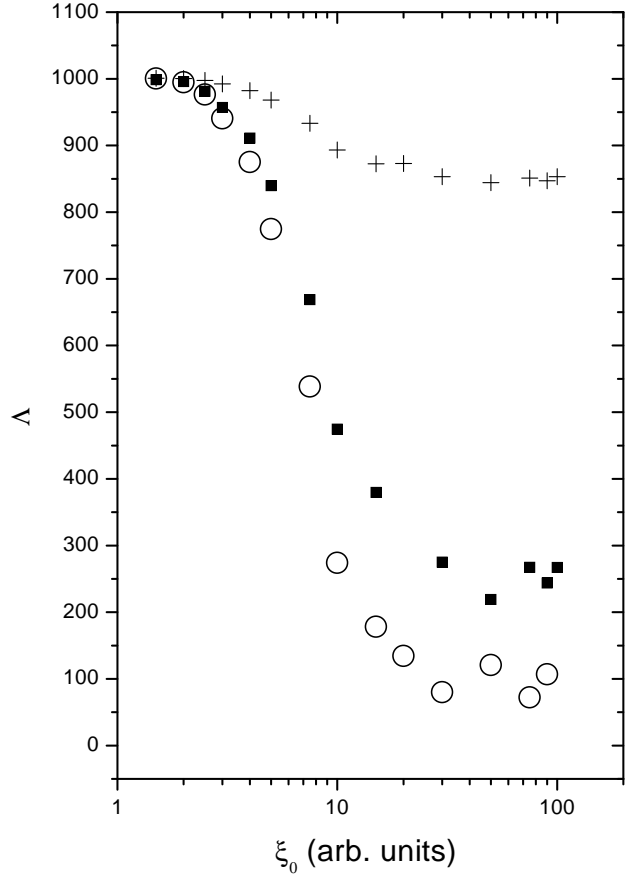}}
\caption{$\Lambda_0$ vs $\xi_0$. The simulation time in this and
the following figures was $t=1000$. The circles represents the
data for $\eta_2 = 0.3$, the squares for $\eta_2 = 0.5$ and the
crosses for $\eta_2 = 0.9$} \label{f54}
\end{figure}

Figure \ref{f54} shows the total length displacement or number of
different lattice sites visited, that we indicate with $\Lambda$,
as a function of the noise intensity $\xi_0$, for different values
of $\eta_2$. For small noise intensity there are no differences
among different values of $\eta_2$, but for increasing $\xi_0$ the
number of visited sites is strongly reduced for $\eta_2=0.3, 0.5$,
but remains still high for $\eta_2=0.9$. This is again consistent
with a larger bias to the origin with $\eta_2$ larger.

\begin{figure}
\centering
\resizebox{.6\columnwidth}{!}{\includegraphics{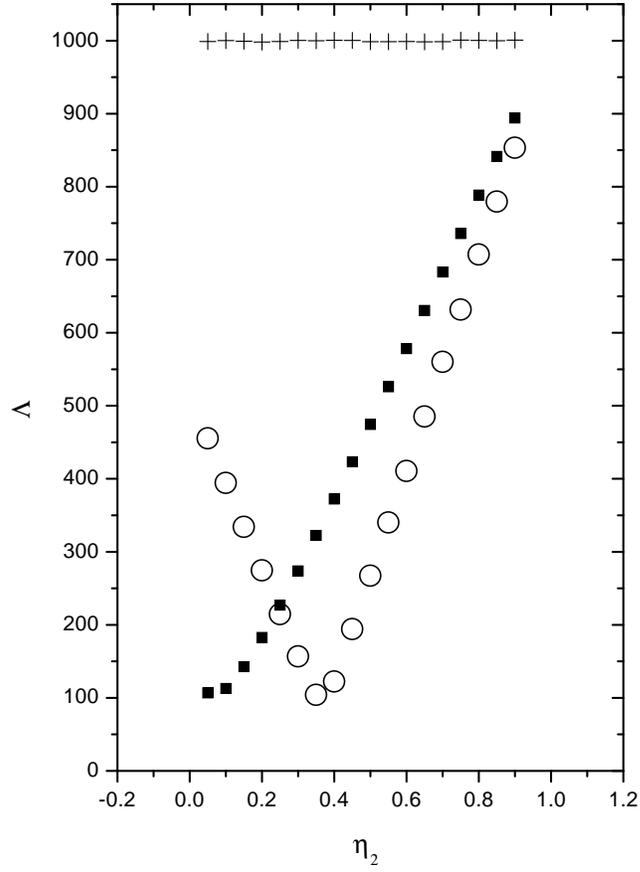}}
\caption{$\Lambda$ vs $\eta_2$. The crosses represents the data
for $\xi_0 = 1.5$, the squares for $\xi_0 = 10.0$ and the circles
for $\xi_0 = 100.0$.} \label{f64}
\end{figure}

\begin{figure}
\centering
\resizebox{.6\columnwidth}{!}{\includegraphics{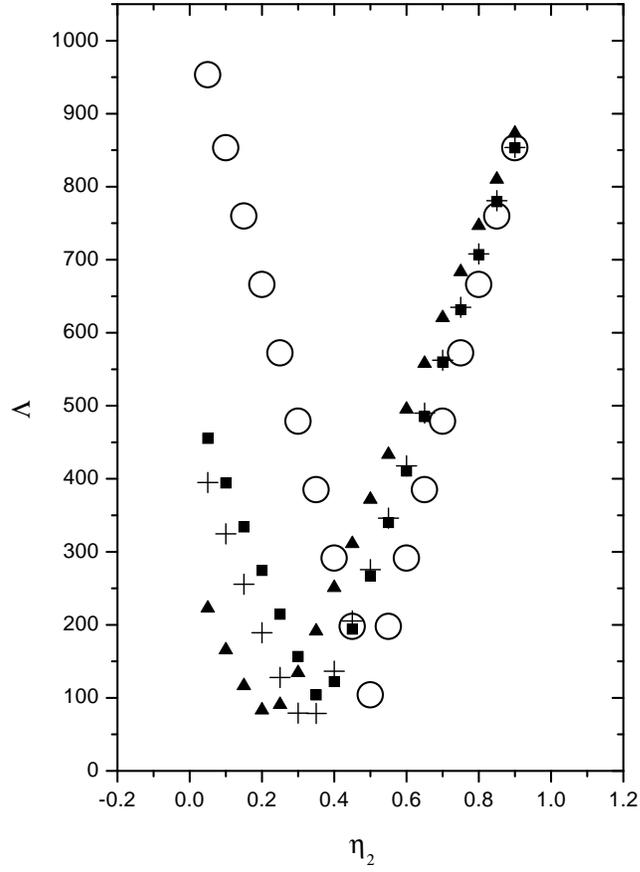}}
\caption{$\Lambda$ vs $\eta_2$.  The result for a one-state
lattice is indicated by circles. The black-squares represents the
data for $\xi_0 = 100.0$, the black-triangles for $\xi_0 = 30.0$
and the crosses for $\xi_0 = 20.0$.} \label{f74}
\end{figure}

\begin{figure}
\centering
\resizebox{.6\columnwidth}{!}{\includegraphics{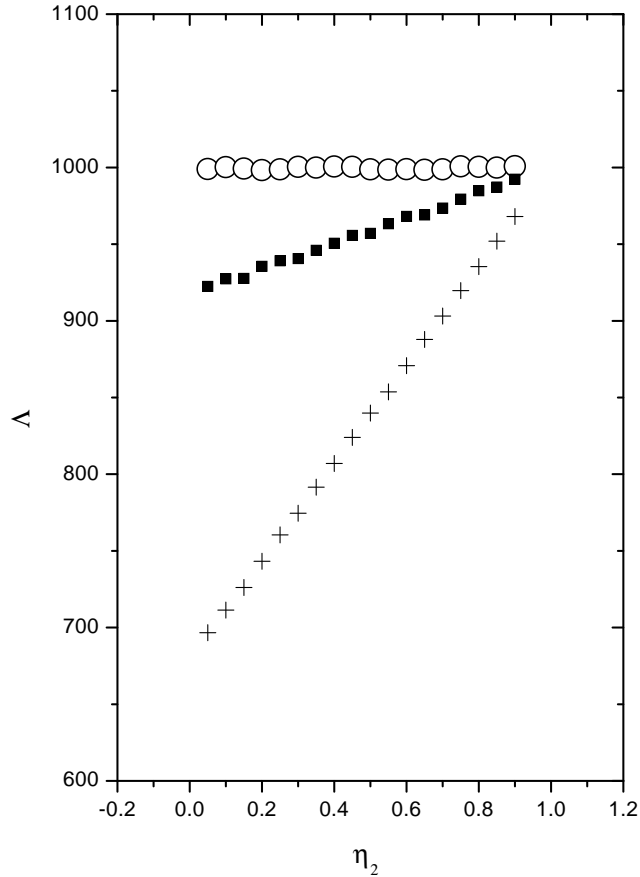}}
\caption{$\Lambda$ vs $\eta_2$. The circles represents the data
for $\xi_0 = 1.5$, the squares for $\xi_0 = 3.0$ and the crosses
for $\xi_0 = 5.0$.} \label{f84}
\end{figure}

Figures \ref{f64}, \ref{f74} and \ref{f84} show the total length
displacement or number of visited lattice sites, indicated with
$\Lambda$, as a function of $\eta_2$ and for different values of
$\xi_0$. In all three cases the simulation time was $t=1000$, and
the initial state of the system was the state $1$. In Fig.
\ref{f64} we compare different cases where the noise intensities
differ by an order of magnitude. When $\xi_0 \simeq 1$ (more
precisely $\xi_0 =1.5$), the number of visited sites remains
almost constant with $\eta_2$. For $\xi_0 = 10$, the curve starts
at a small number of sites and increases monotonically, almost
linearly, with $\eta_2$. For larger values of $\xi_0$ ($\xi_0 =
100$), the presence of a minimum is apparent. A remarkable fact is
that before and after the minimum in $\Lambda$, $\Lambda$ behaves
linearly with $\eta_2$. It seems that the high number of visited
sites for small $\xi_0$ when $\eta_2$ is also small, reduces with
increasing $\xi_0$ before rising again for still larger values of
$\xi_0$. This argument was supported by complementary simulations,
see for instance Fig. \ref{f74} where we have depicted the same
situation but now for a lattice with only one state, as well as
for the two-state lattice and in the range of large noise
intensities ($\xi_0 = 20, 30, 100$). The indicated effect is
apparent. For the one-state lattice, the minimum is located at
$\eta = 0.5$. The reason is that $\eta = 0.5$ corresponds to the
situation of unbiased particle motion, that is we have the same
probability of moving to the left and to the right of the lattice.
This is a point of symmetry in the sense that a larger or smaller
$\eta$ implies a biased motion in one direction or the other
(changing $\eta$ by $1-\eta$ only changes the direction of motion
but not the number of visited sites).

We can see how the noise changes the position of the minimum in
the two-state lattice relative to the one-state case. The strong
dependence of the system $'s$ response for small bias is apparent,
as well as the linear scaling of $\Lambda$ with $\eta_2$ before
and after the minimum . The decrease in the value for small
$\eta_2$ is also shown in Fig. \ref{f84}, where we have depicted
the same situation as before but now for $\xi_0 = 1.5, 3.$ and
$5.$. The indicated decrease is clearly seen. From this figure we
can see that the indicated minimum disappears for small noise
intensities, and the ``full" linear dependence of $\Lambda$ on
$\eta_2$ in this range.

\subsection{Conclusions}

We have studied here the evolution of particles diffusing in a
fluctuating medium when the switching mechanism that governs the
transition between the states has two components: one deterministic
and the other stochastic. We have chosen a periodic signal for the
deterministic part while the stochastic component is characterized
by a Gaussian white noise of zero mean. The parameters were chosen
in such a way that the effect of the deterministic mechanism alone
is not enough to produce the switching, but the simultaneous
action of the deterministic and the stochastic signals can induce
the jump over the potential barrier and produce the switching. The
problem was characterized studying the MFPT in an infinite system,
and we have also analyzed the probability that the particle
returns to the initial position, and the total length displacement
or number of visited lattice sites.

The main goal of this work is the observation (see fig. \ref{f14})
of a {\it resonant activation}-like phenomenon when considering
the MFPT as a function of the noise intensity of the stochastic
component of the transition rate. Another important result is the
kind of {\it double resonant} behavior, both in $\xi_0$ and
$\eta_2$, observed in fig. \ref{f34}.

It is also worth remarking here the behavior of the number of
visited lattice sites ($\Lambda$) as a function of $\eta_2$, with
$\xi_0$ as a parameter. For the case of the one-state lattice, a
minimum in the number of visited lattice sites exists for $\eta =
0.5$. For the two-state lattice, that minimum changes with the
noise intensity. For large values of $\xi_0$, it shifts but is
still well marked. However, when $\xi_0$ is reduced the minimum is
less marked and finally, for very small values of $\xi_0$, it
disappears. This is another example of the ``constructive" role
that noise can play, in this case associated to diffusion on a
lattice with several (here two) states. Another remarkable fact is
the ``piece-wise" linear dependence of $\Lambda$ on $\eta_2$

According to the discussion above on the motivations of the
present work, a natural step in this research of diffusion in a
fluctuating medium will be, as in Ref. \cite{I122}, the inclusion
of a gated trap mechanism at the origin. The aim will be to
analyze the interplay between both deterministic-plus-stochastic
mechanisms, the switching and the gating one, and the coherent or
interference effects that could arise. This problem will be the
subject of further work.

\vspace{0.25cm}

{\bf Acknowledgments:} The authors thank V. Gr\"unfeld for a
critical reading of the manuscript. HSW acknowledges the partial
support from ANPCyT, Argentine, and thanks the MECyD, Spain, for
an award within the {\it Sabbatical Program for Visiting
Professors}, and to the Universitat de les Illes Balears for the
kind hospitality extended to him.


\end{document}